\documentclass[12pt,dvips]{article}
\usepackage{graphics}
\def\lsim{\mathrel{\lower2.5pt\vbox{\lineskip=0pt\baselineskip=0pt
           \hbox{$<$}\hbox{$\sim$}}}}
\def\gsim{\mathrel{\lower2.5pt\vbox{\lineskip=0pt\baselineskip=0pt
           \hbox{$>$}\hbox{$\sim$}}}}

\def\beq{\begin{equation}}
\def\eeq{\end{equation}}
\def\bea{\begin{eqnarray}}
\def\eea{\end{eqnarray}}
\def\nn{\nonumber}
\def\dfrac{\displaystyle\frac}

\def\eqref#1{Eq.(\ref{eqn:#1})}

\def\eqlab#1{\label{eqn:#1}}

\def\l{\left}
\def\r{\right}

\def\Su{\rm u}
\def\Sd{\rm d}
\def\Sud{\rm u,d}

\def\tr{{\rm tr}}

\def\hu{{y^{\rm u}}}
\def\hd{{y^{\rm d}}}
\def\he{{y^{\rm e}}}
\def\hn{{y^{\nu}}}

\begin{document}
\setlength{\baselineskip}{8mm}
\begin{titlepage}
\begin{flushright}
\begin{tabular}{c c}
& {\normalsize  hep-ph/9810471} \\
& {\normalsize  DPNU-98-36} \\
& {\normalsize  KEK-TH-597} \\
& {\normalsize  OHSTPY-HEP-T-98-022} \\
& {\normalsize Oct. 1998}
\end{tabular}
\end{flushright}
\vspace{5mm}
\begin{center}
{\large The renormalization group analysis of
 the large lepton flavor mixing and the neutrino mass
}\\

\vspace{15mm} 
{N.~Haba}$^1$,\footnote{E-mail: haba@pacific.mps.ohio-state.edu} 
\hspace{2mm}
{N.~Okamura}$^2$,\footnote{E-mail: okamura@theory.kek.jp} 
\hspace{2mm}
{M.~Sugiura}$^3$ \footnote{E-mail: sugiura@eken.phys.nagoya-u.ac.jp} \\

\vspace{5mm}
$^1${\it Department of Physics, The Ohio State University,
Columbus, Ohio 43210, USA
} \\
$^2${\it Theory Group, KEK, Oho 1-1, Tsukuba 305-0801, Japan
} \\
$^3${\it Department of Physics, Nagoya University, 
Nagoya, 464-8602, Japan
} \\

\end{center}

\vspace{10mm}


\begin{abstract}

The Superkamiokande experiment suggests 
the large flavor mixing between 
$\nu_{\mu}$ and $\nu_{\tau}$. 
We show that 
 the mixing angle receives significant corrections
 from the 
 renormalization group equation (RGE) 
 when both the second and the third generation neutrino masses 
 are larger than $O(0.1\mbox{eV})$. 
This means that 
 the mixing angle must be small
 at the decoupling scale of right-handed neutrinos 
 in the model containing a sterile neutrino $\nu_s$
 with the mass spectrum of 
 $m_{\nu_s} \approx m_{\nu_e} \ll 
  m_{\nu_{\mu}} \approx m_{\nu_{\tau}}$.


\end{abstract}
\end{titlepage}

%
\section{Introduction}

The recent Superkamiokande data suggests the 
large neutrino flavor mixing between 
$\nu_{\mu}$ and $\nu_{\tau}$\cite{SK1}. 
According to this experimental result,
there have been a lot of theoretical attempts
to explain why 
the large flavor mixing is realized in 
the lepton sector\cite{ex1}. 
One of the interesting approaches 
is concentrating on the effects 
of the the renormalization group 
equation (RGE). 
The RGE effects
cause the enhancement of the neutrino flavor
mixing in some situations\cite{Babu}\cite{Tanimoto}\cite{MHE}.

\par
In this paper we analyze the RGE 
of the neutrino flavor mixing 
between $\nu_{\mu}$ and $\nu_{\tau}$
in the minimal supersymmetric standard model (MSSM) 
with right-handed neutrinos. 
%
%
Here we stand the position that 
the smallness of neutrino masses is explained 
 by the seesaw mechanism\cite{seesaw}.
We consider the situation that 
 $m_{\nu_e}$
 is much smaller than $m_{\nu_{\mu}}$ and $m_{\nu_{\tau}}$, 
 and expect the solar neutrino problem\cite{2}
 is solved by the oscillation between
 $\nu_e$ and a sterile neutrino $\nu_s$ \cite{Hata}.
This situation is so-called
 four light neutrino scenario\cite{4Neutrino}\cite{4NeutrinoGUT} 
\footnote{
The four neutrino scenario might also explain 
 the results of the LSND\cite{LSND}. 
The LSND results suggest the small mixing
 between $\overline{\nu}_{\mu}$ and
 $\overline{\nu}_{e}$ with 
 $m^2_{\nu_{\mu}} - m^2_{\nu_{e}} \sim 1\mbox{eV}$. 
However, the confirmation of this 
 result still awaits future experiments.
Recent measurements in the KARMEN
 detector exclude part of the 
 LSND allowed region\cite{KARMEN}.
}  with the mass spectrum of  
 $m_{\nu_s} \approx m_{\nu_e} \ll 
  m_{\nu_{\mu}} \approx m_{\nu_{\tau}}$.
In this neutrino mass hierarchy, 
 we find that 
 the mixing angle between $\nu_{\mu}$ 
 and $\nu_{\tau}$  
 receives significant corrections
 from the 
 renormalization group equation (RGE). 
In this case 
 the mixing angle at the high energy  
 must be small 
 as long as the mixing at the low energy 
 is maximal. 
{}From the view point of the model building, 
 we must find the fundamental theory which induces
 the small mixing angle at the high energy 
 in the four neutrino scenario with the 
 above mass spectrum.



\par
\section{The RGE effects of the neutrino flavor mixing}

\subsection{The RGEs of the Yukawa couplings}
In this section
 we show the RGEs of the MSSM with right-handed neutrinos.
The superpotential of the MSSM is given by
\beq
{\cal W}
   = \hu_{ij} {Q_i} H_{\Su} {\bar U_j} 
     + \hd_{ij} {Q_i} H_{\Sd} {\bar D_j}
     + \hn_{ij} {L_i} H_{\Su} {\bar N_j}
     + \he_{ij} {L_i} H_{\Sd} {\bar E_j}
     + \mu_H H_{\Su} H_{\Sd}
     + \frac12 M_{R\;ij} {\bar N_{i}} {\bar N_{j}},
\label{SuperPotMSSMH}
\eeq
 where the index $i,j$ stands for the generation 
number $(i,j = 2,3)$.
In this paper we neglect Yukawa couplings
of the first generation since
we consider the case where
$m_{\nu_e}$ is much smaller than 
$m_{\nu_{\mu}}$ and $m_{\nu_{\tau}}$.
$Q_i$, $L_i$, ${\bar U}_i$, ${\bar D}_i$, ${\bar E}_i$, ${\bar N}_i$
 and $H_{\Sud}$ are quark doublet, lepton doublet, 
 right-handed up-sector, right-handed down-sector, 
 right-handed charged lepton, right-handed neutrino
 and Higgs fields, respectively. 
$M_{R\;ij}$ is the Majorana mass matrix of 
the right-handed neutrinos, which is symmetric
 under the generation indices $i,j$. 
$\mu_H$ is the supersymmetric mass parameter of Higgs particles. 
\par

In this model the RGEs of Yukawa couplings are given by
\bea
\dfrac{d}{dt}\hu & = & \dfrac{1}{\l( 4\pi \r)^2}
     \l[\tr \l( 3 \hu \hu^{\dagger} + \hn \hn^{\dagger} \r)
        + 3 \hu \hu^{\dagger} + \hd \hd^{\dagger} 
     -4\pi\l(\dfrac{16}{3}\alpha_3+3\alpha_2+\dfrac{13}{15} \alpha_1\r)
     \r] \hu , 
\nn
\\
\dfrac{d}{dt}\hd & = & \dfrac{1}{\l( 4\pi \r)^2}
      \l[\tr \l( 3 \hd \hd^{\dagger} + \he \he^{\dagger} \r)
       + 3 \hd \hd^{\dagger} + \hu \hu^{\dagger} 
     -4\pi\l(\dfrac{16}{3}\alpha_3+3\alpha_2+\dfrac{7}{15} \alpha_1\r)
     \r] \hd, 
\nn
\\
\dfrac{d}{dt}\he & = & \dfrac{1}{\l(4\pi\r)^2}
      \l[\tr \l( 3 \hd \hd^{\dagger} + \he \he^{\dagger} \r)
       + 3 \he \he^{\dagger} + \hn \hn^{\dagger}
     -4\pi\l(3\alpha_2+\dfrac{9}{5} \alpha_1\r) \r] \he,
\nn
\\
\dfrac{d}{dt}\hn & = & \dfrac{1}{\l(4\pi\r)^2}
      \l[\tr \l( 3 \hu \hu^{\dagger} + \hn \hn^{\dagger} \r)
       + 3 \hn \hn^{\dagger} + \he \he^{\dagger} 
     -4\pi\l(3\alpha_2+\dfrac{3}{5} \alpha_1\r)\r] \hn,
\eqlab{RGE-high}
\eea
 where $t = \ln \mu$ and $\mu$ is a renormalization point, respectively.
These equations are available for the energy region of 
$\mu > {\cal M_R}$,
where ${\cal M_R}$ denotes
the energy scale of the Majorana mass.
\par

Below the scale of ${\cal M_R}$, 
we should take the decoupling effects 
of heavy neutrinos into account.
The effective theory is described without heavy neutrinos.
The superpotential of Eq.(\ref{SuperPotMSSMH}) is modified as
\beq
{\cal W}
   = \hu_{ij} {Q_i} H_{\Su} {\bar U_j} 
     + \hd_{ij} {Q_i} H_{\Sd} {\bar D_j}
     + \he_{ij} {L_i} H_{\Sd} {\bar E_j}
     - \frac12 \kappa_{ij}{\nu_{i}} {\nu_{j}} H_{\Su} H_{\Su}.
\label{SuperPotMSSML}
\eeq
Here $\nu_i$s are the light modes of neutrinos 
 which remain after integrating out the heavy ones.
The coupling constant 
$\kappa_{ij}$ is defined as
\beq
  \kappa_{ij} = ( \hn \; M_R^{-1}\; \hn^{T})_{ij}\; .
\eeq
It relates to the mass matrix of the light neutrinos as
\newcommand{\VEV}[1]{{\langle {#1} \rangle}}
\beq
  m^\nu_{ij} = \dfrac{v_{\Su}^2}{2} \kappa_{ij}
             = \dfrac{v^2 \sin^2 \beta}{2}  \kappa_{ij}\; ,
  \eqlab{NeutrinoMass} 
\eeq
where 
\beq
\tan\beta \equiv \dfrac{v_{\Su}}{v_{\Sd}},~~~~~~~
 v^2 = v_{\Su}^2 + v_{\Sd}^2,
\eeq
 with $\langle H_{\Su}\rangle =v_{\Su}$ and 
 $\langle H_{\Sd}\rangle = v_{\Sd}$. 
The value of $v$ is given by 
\beq
 v = M_Z \dfrac{\sin{2\theta_W}}{2}
         \sqrt{\dfrac{\alpha}{\pi}} = 245.4 \l(\mbox{GeV}\r),
 \eqlab{vev}
\eeq
 with $M_Z=91.187$ GeV,
 $\alpha=127.9$, 
 and $\sin^2{\theta_W}=0.23$\cite{PDG}.
\par

In $\mu < {\cal M_R}$,
 the RGEs of the Yukawa couplings
 Eqs.(\ref{eqn:RGE-high}) are modified as 
\bea
\dfrac{d}{dt}\hu & = & \dfrac{1}{\l(4\pi\r)^2}
     \l[\tr\l(3\hu \hu^{\dagger}\r) + 
     3\hu\hu^{\dagger} + \hd\hd^{\dagger}
    -4\pi\l(\dfrac{16}{3}\alpha_3+3\alpha_2+\dfrac{13}{15}\alpha_1\r)
      \r] \hu, 
\nn
\\
\dfrac{d}{dt}\hd & = & \dfrac{1}{\l(4\pi\r)^2}
      \l[\tr\l(3\hd\hd^{\dagger}+\he\he^{\dagger}\r)
       + 3\hd\hd^{\dagger} + \hu\hu^{\dagger}
     -4\pi\l(\dfrac{16}{3}\alpha_3+3\alpha_2+\dfrac{7}{15}\alpha_1\r) \r] \hd,
\nn
\\
\dfrac{d}{dt}\he & = & \dfrac{1}{\l(4\pi\r)^2}
      \l[\tr\l(3\hd\hd^{\dagger}+\he\he^{\dagger}\r)
       + 3\he\he^{\dagger}
     -4\pi\l(3\alpha_2+\dfrac{9}{5} \alpha_1\r) \r] \he,
\eqlab{RGE-low}
\eea
and 
\beq
 \dfrac{d}{dt}\kappa = {1 \over 8\pi^2}
    \l[
      \l\{ \tr \l( 3\hu\hu^{\dagger} \r) 
       - 4\pi \l( 3\alpha_2 + \dfrac{3}{5} \alpha_1 \r) \r\} \kappa
+\dfrac{1}{2} \l\{ \l( \he \he^{\dagger}\r) \kappa
                 + \kappa\l( \he \he^{\dagger}\r)^{T} \r\}
    \r].
\eqlab{RGE_kappa}
\eeq
{}From Eqs.(\ref{eqn:RGE-low}), we can see that 
 the RGEs of quark and charged lepton
 do not include the neutrino Yukawa couplings 
 contrary to the case of $\mu >{\cal M_R}$ as Eqs.(\ref{eqn:RGE-high}). 
Hence below ${\cal M_R}$ the running of the Yukawa couplings
 of quark and charged lepton can be determined
 independently of that of the neutrinos.
%

\subsection{The RGEs of neutrinos in the effective theory}
{}From now on we are concentrating on 
 the RGE effects below the scale of ${\cal M_R}$,
 which are given by Eqs.(\ref{eqn:RGE-low}) and (\ref{eqn:RGE_kappa}). 
Since RGEs of the Yukawa couplings for quark and charged lepton of
 Eqs.(\ref{eqn:RGE-low}) 
 can be solved without information about the neutrino sector
 as mentioned before,
 the renormalization point dependences of $y_\mu$ and $y_\tau$
 are completely determined
 by the RGEs' boundary conditions which we take 
 the masses of quark and charged lepton\footnote{
We use $m_t = 174.5$ GeV, $m_c = 0.657$ GeV, 
$m_b = 3.02$ GeV, $m_s = 9.935 \times 10^{-2}$ GeV, 
$m_{\tau} = 1.746$ GeV, and $m_{\mu} = 1.0273 \times 10^{-3}$ GeV
at $\mu = M_Z$\cite{PDG}. 
},
 the Cabibbo-Kobayashi-Maskawa matrix \cite{CKM}
 and $\tan\beta$ at the weak scale. 
Then we have only to concentrate on Eq.(\ref{eqn:RGE_kappa}).
Here we neglect $CP$ phases in the flavor mixing matrices
 of the quark and the lepton sector for simplicity. 
\par

{}For convenience, 
 we take three independent parameters 
 $\kappa_r \equiv \kappa_{22}/\kappa_{33}$,
 $\sin^2 2\theta_{23}$ and $\delta \kappa^2$ 
 instead of $\kappa_{ij}$ 
 [$\kappa_{22}$, $\kappa_{33}$ and 
  $\kappa_{23} (= \kappa_{32})$]. 
Here $\sin\theta_{23}$ and $\delta \kappa^2$
 are determined from $\kappa_{ij}$ by the following equations:
\bea
  \kappa &=&
    \left(
      \begin{array}{cc}
          \cos \theta_{23} & \sin \theta_{23} \\
        - \sin \theta_{23} & \cos \theta_{23}
      \end{array}
    \right)
    \left(
      \begin{array}{cc}
        \kappa_2 & 0 \\
        0 & \kappa_3
      \end{array}
    \right)
    \left(
      \begin{array}{cc}
        \cos \theta_{23} & - \sin \theta_{23} \\
        \sin \theta_{23} &   \cos \theta_{23}
      \end{array}
    \right),
    \eqlab{KappaToMixing} \\
  \delta \kappa^2 &\equiv&
    \kappa_3^2 - \kappa_2^2,
    \eqlab{DeltaKappa}
\eea
 where 
\bea
    \kappa_3 =
    \dfrac{\sqrt{\delta \kappa^2}}{2}
    \left(
      \sqrt{\alpha} + \dfrac{1}{\sqrt{\alpha}}
    \right), \;\;\;
    \kappa_2 =
    \dfrac{\sqrt{\delta \kappa^2}}{2}
    \left(
      \sqrt{\alpha} - \dfrac{1}{\sqrt{\alpha}}
    \right),
  \eqlab{EVal2KrSin}
\eea
 with
\beq
   \alpha \equiv
   \left|
     {\dfrac{1 + \kappa_r}{1 - \kappa_r}} \cos 2\theta_{23}
   \right| .
\eeq
By using this relation, 
 the RGE of Eq.(\ref{eqn:RGE_kappa}) can be rewritten into
 the following three equations\footnote{
Equation (\ref{eqn:RGEkappa2b}) was firstly 
 derived by Babu, Leung and Pantaleone
 in Ref.~\cite{Babu}.
}:
\bea\
  \dfrac{d}{dt} \kappa_r &=&
      - \dfrac{1}{8\pi^2}(y_\tau^2 - y_\mu^2)\kappa_r,
  \eqlab{RGEkappa2a}\\
  \dfrac{d}{dt} \sin^2 2\theta_{23} &=&
      - \dfrac{1}{8\pi^2}\sin^22\theta_{23}(1-\sin^22\theta_{23})
      (y_\tau^2 - y_\mu^2)
      \dfrac{1 + \kappa_r}{1 - \kappa_r},
  \eqlab{RGEkappa2b}\\
  \dfrac{d}{dt} \delta \kappa^2 &=&
      \dfrac{1}{8\pi^2}
      \Bigg[
        2 \left\{
          \tr \l( 3\hu\hu^{\dagger} \r) 
          - 4\pi \l( 3\alpha_2 + \dfrac{3}{5} \alpha_1 \r)
        \right\}
        + y_\tau^2 + y_\mu^2 \nonumber \\
 & & +  \left( y_\tau^2 - y_\mu^2 \right)
        \left(
          \dfrac{1 + \kappa_r^2}{1 - \kappa_r^2}
          - \frac12 \cdot 
          \dfrac{1 + \kappa_r}{1 - \kappa_r} \sin^2 2 \theta_{23}
        \right)
      \Bigg] \delta \kappa^2.
\eqlab{RGEkappa2c}
\eea
Both $\sin^2 2\theta_{23}$ and $\delta \kappa^2$ 
 directly relate to the observed quantities
 in neutrino oscillation experiments%
\footnote{%
Since we take a diagonal base of the charged lepton,
$\sin^2 2\theta_{23}$ of this paper 
is an observable quantity. 
}.
The mass squared difference can be written as
 $\delta m^2 = v^4\sin^4\!\beta\; \delta\kappa^2/4$
 by using Eq.(\ref{eqn:NeutrinoMass}).
\par

As shown before, 
 $y_\mu(\mu)$ and $y_\tau(\mu)$ are determined 
 without knowing the neutrino Yukawa coupling,
 thus we can obtain values of 
 $\sin 2 \theta_{23}$ and $\kappa_r$ 
 at the weak scale by using only 
 Eqs.(\ref{eqn:RGEkappa2a}) and (\ref{eqn:RGEkappa2b}). 
We analyze RGEs of Eqs.(\ref{eqn:RGEkappa2a}) and (\ref{eqn:RGEkappa2b})
 by inputting various values of $\sin \theta_{23}$ and $\kappa_r$ for 
 the initial conditions at $\mu ={\cal M_R}$\footnote{
As shown later, we do not need to
 calculate Eq.(\ref{eqn:RGEkappa2c}) in our analysis.
}.

\subsection{Numerical results of the RGEs}

\begin{figure}[htbp]
 \begin{center}
 \resizebox{.8\textwidth}{!}{\includegraphics{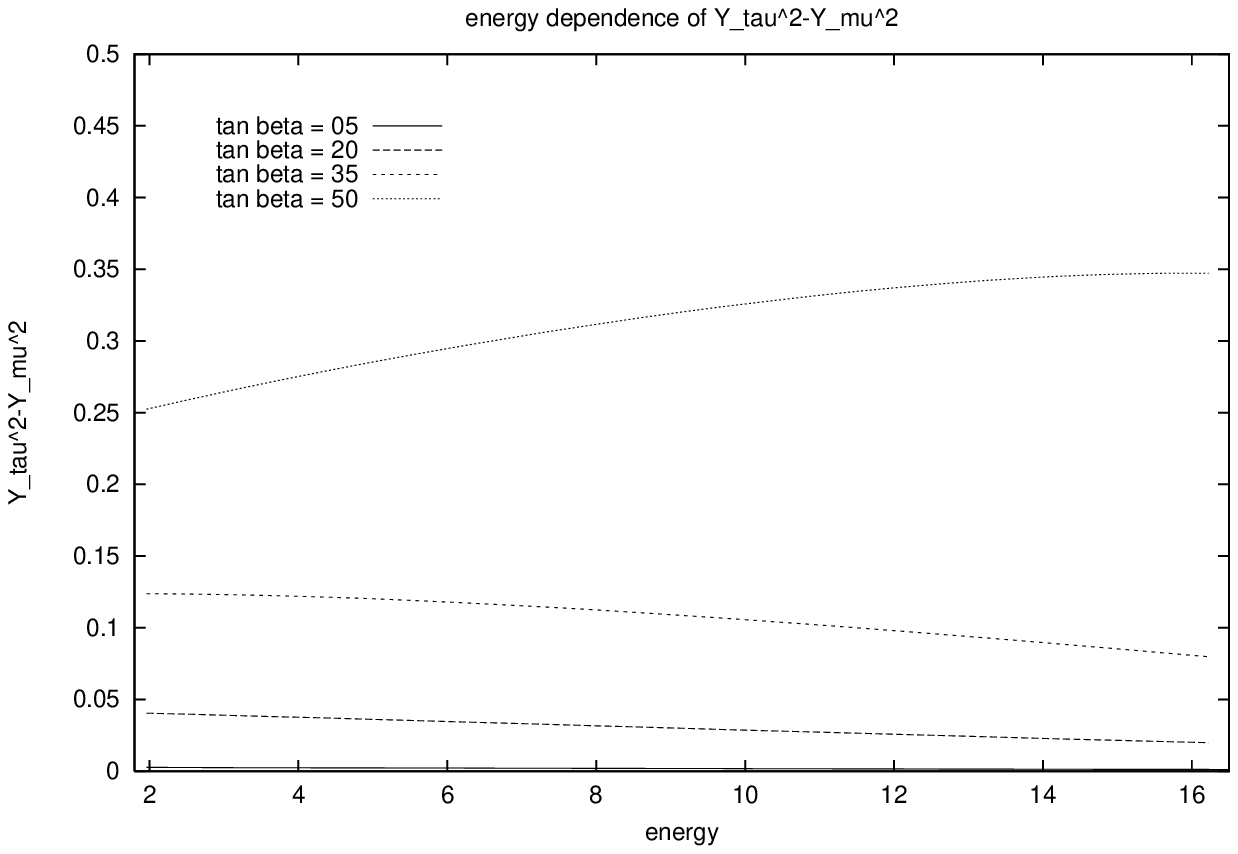}}
 \end{center}
 \vspace{-1em}
 \caption{
The energy dependence of the values of 
$y_\tau^2 - y_\mu^2$ with various values of 
$\tan\beta$. 
These values are the coefficients of the RGEs of 
$\kappa_r$ and 
$\sin^2 2 \theta_{23}$. 
}
 \label{fig:Yem}
\end{figure}

Now we show the numerical results of the RGEs. 
{}Figure \ref{fig:Yem} shows the 
 energy dependence of the values of $[y_\tau^2 - y_\mu^2]$,
 which is the coefficients of the RGEs of 
 $\kappa_r$ in \eqref{RGEkappa2a} and $\sin^2 2 \theta_{23}$
 in \eqref{RGEkappa2b}.
{}Four lines
 correspond to the various values of 
 $\tan\beta$, which we take 5, 20, 35, and 50.
The values of $[y_\tau^2 - y_\mu^2]$ do not receive 
 the significant RGE corrections when $\tan \beta$ is small. 
\par

\begin{figure}[htbp]
 \begin{center}
 \resizebox{.75\textwidth}{!}{\includegraphics{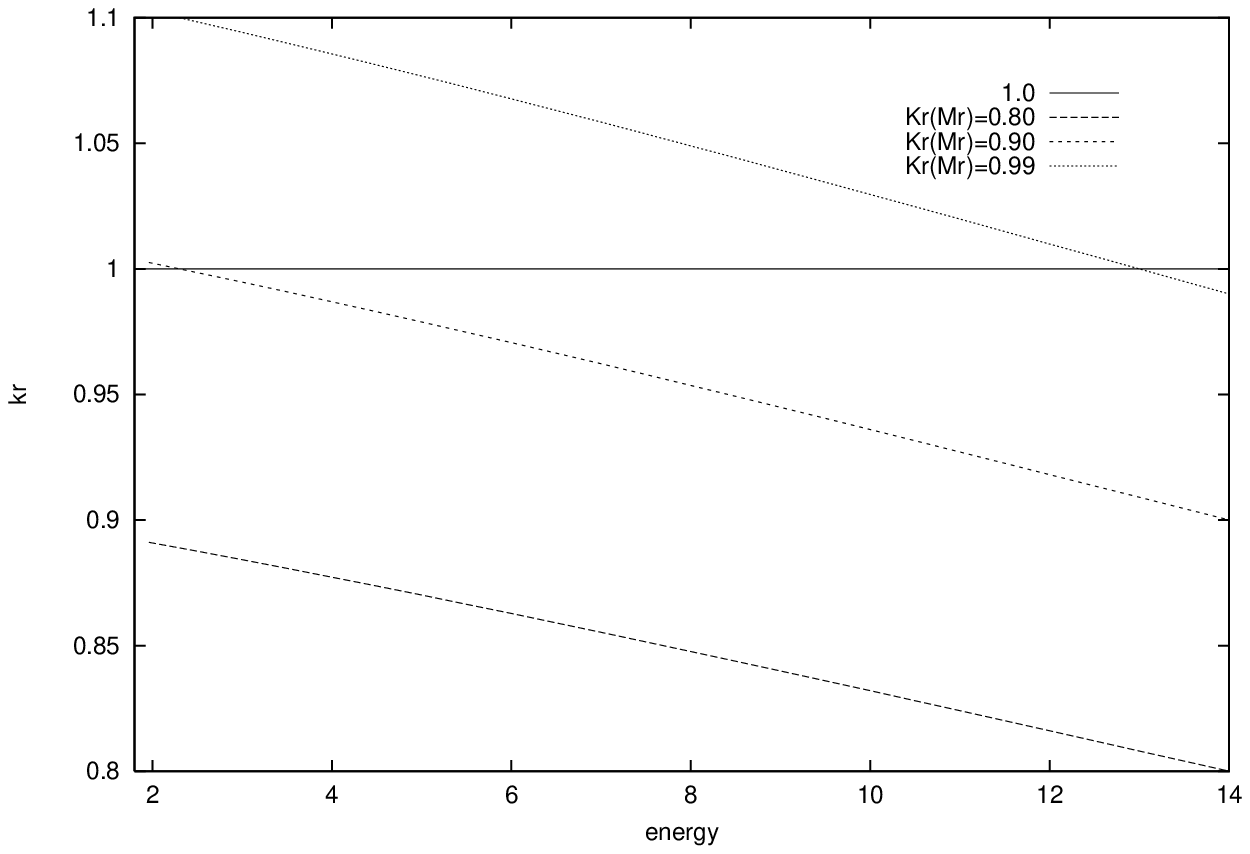}}\\
 Figure~\ref{fig:Kr}(a): ${\cal M_R}=10^{14}$ GeV \\
 \vspace{1em}\par\noindent
 \resizebox{.75\textwidth}{!}{\includegraphics{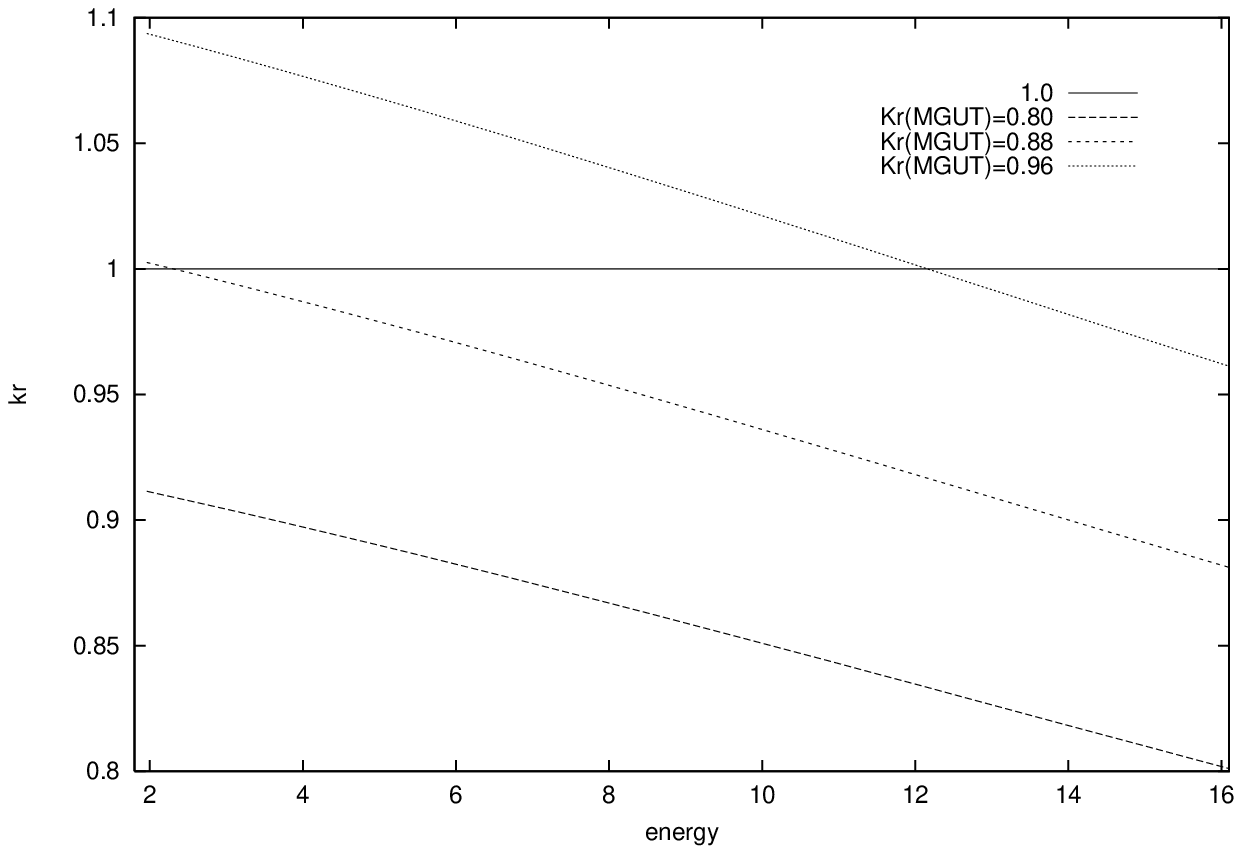}}\\
 Figure~\ref{fig:Kr}(b): ${\cal M_R}=10^{16}$ GeV
 \end{center}
 \vspace{-1em}
 \caption{
The energy dependence of $\kappa_r$ with 
$\tan\beta =50$. 
(a): The initial condition is fixed as 
$\kappa_r({\cal M_R}) = 0.8, 0.9$, and $0.99$ 
at ${\cal M_R}=10^{14}$ GeV.
(b): The initial condition is fixed as 
$\kappa_r({\cal M_R}) = 0.8, 0.88$, and $0.96$ 
at ${\cal M_R}=10^{16}$ GeV. 
}
 \label{fig:Kr}
\end{figure}

{}Figures \ref{fig:Kr}
 show the energy dependence of $\kappa_r$ with $\tan\beta = 50$.
We show 
 the two cases of (a):${\cal M_R}= 10^{14}$ GeV and 
 (b):$10^{16}$ GeV 
 with various initial conditions at ${\cal M_R}$.
{}From the high energy to the low energy,
$\kappa_r$ simply increases.
If we take the small $\tan \beta$, 
 the slope tends to be flat because
 the value of $[y_{\tau}^2-y_\mu^2]$ in \eqref{RGEkappa2a}
 decreases as Fig.~\ref{fig:Yem}.
\par

\begin{figure}[htbp]
 \begin{center}
  \resizebox{.75\textwidth}{!}{\includegraphics{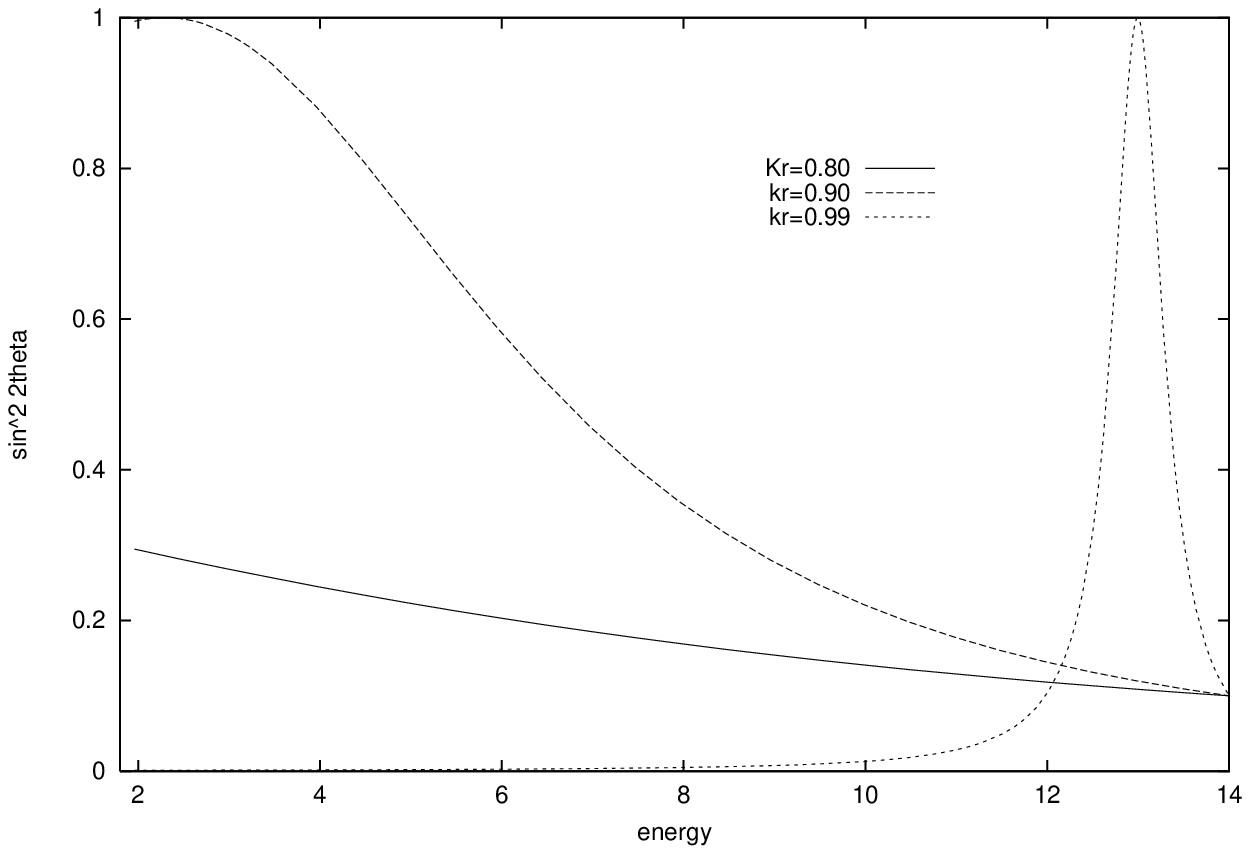}}\\
  Figure~\ref{fig:EngDepOfSin}(a): ${\cal M_R}=10^{14}$ GeV \\
  \vspace{1em}\par\noindent
  \resizebox{.75\textwidth}{!}{\includegraphics{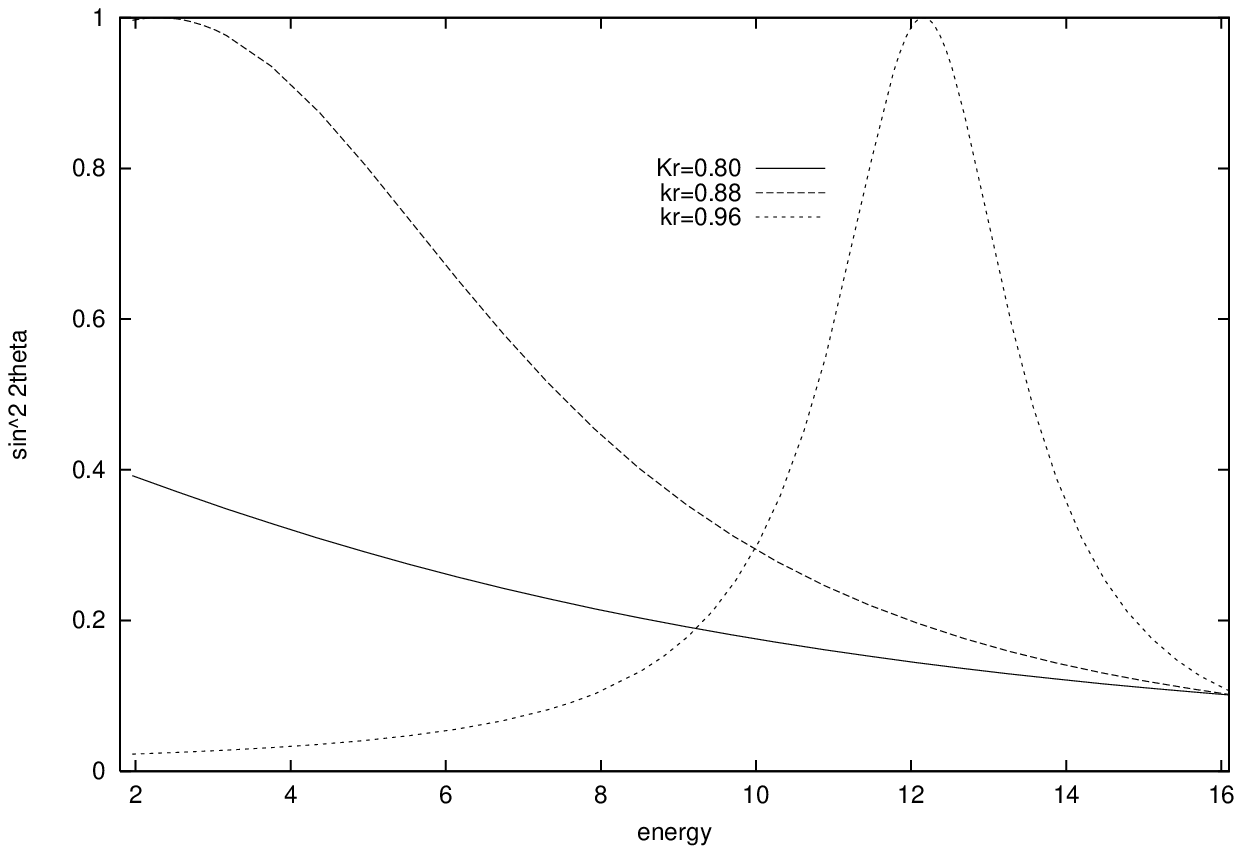}}\\
  Figure~\ref{fig:EngDepOfSin}(b): ${\cal M_R}=10^{16}$ GeV
 \end{center}
\caption{%
The energy dependence of $\sin^2 2 \theta$ 
with $\tan \beta =50$. 
We show two cases of (a): 
${\cal M_R}=10^{14}$ GeV and 
 (b): $10^{16}$ GeV. 
}
 \label{fig:EngDepOfSin}
\end{figure}

Figures~\ref{fig:EngDepOfSin}
 show the energy dependence of the mixing angle
 with the same values of ${\cal M_R}$ and $\kappa_r({\cal M_R})$
 as Figs.~\ref{fig:Kr}.
There are three curves corresponding to
 the values of $\kappa_r({\cal M_R})$,
 and all of them have the same boundary condition
 of $\sin^2 2\theta_{23}({\cal M_R}) = 0.1$.
{}From these figures we can see that
 the mixing angle at the weak scale changes
 depending on the value of $\kappa_r({\cal M_R})$.
This difference can be easily understood
 by comparing Figs.~\ref{fig:EngDepOfSin} with Figs.~\ref{fig:Kr}
 and by the existence of the factor $[(1+ \kappa_r)/(1- \kappa_r)]$
 in the R.H.S. of the RGE of Eq.(\ref{eqn:RGEkappa2b}).
Let us see Fig.~\ref{fig:Kr}(a) and Fig.~\ref{fig:EngDepOfSin}(a),
 for example.
In the case of $\kappa_r({\cal M_R}) = 0.8$,
 where $\kappa_r$ does not exceed one at all the energy scales,
 the mixing angle does not receive the significant RGE corrections.
It is because the R.H.S. of Eq.(\ref{eqn:RGEkappa2b})
 does not become so large as to enhance the mixing angle.
On the other hand,
 in the case of $\kappa_r({\cal M_R}) = 0.9$
 $\kappa_r$ exceeds one near the weak scale,
 which results in the significant enhancement
 of the mixing angle by the factor $[(1+ \kappa_r)/(1- \kappa_r)]$.
Finally in the case of $\kappa_r({\cal M_R}) = 0.99$,
 $\kappa_r$ exceeds one above the weak scale.
Then the mixing angle once becomes maximal at high energy,
 however, after there it decreases rapidly
 since the sign of $[(1+ \kappa_r)/(1- \kappa_r)]$ changes.

%
\par

\begin{figure}[htbp]
 \begin{center}
 \resizebox{.75\textwidth}{!}{\includegraphics{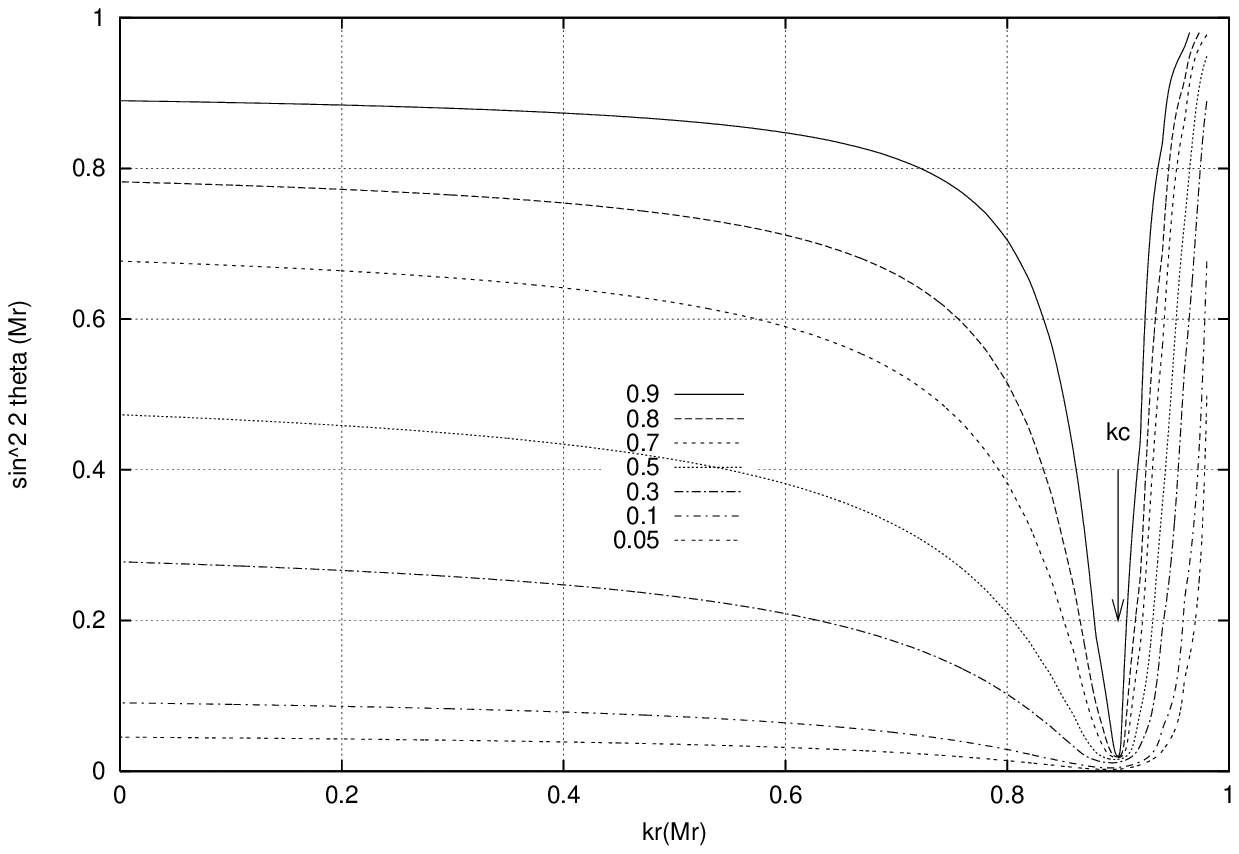}}\\
 Figure~\ref{fig:sin}(a): $\tan\beta=50$, ${\cal M_R}=10^{14}$ GeV\\
 \vspace{.5em}\par\noindent
 \resizebox{.75\textwidth}{!}{\includegraphics{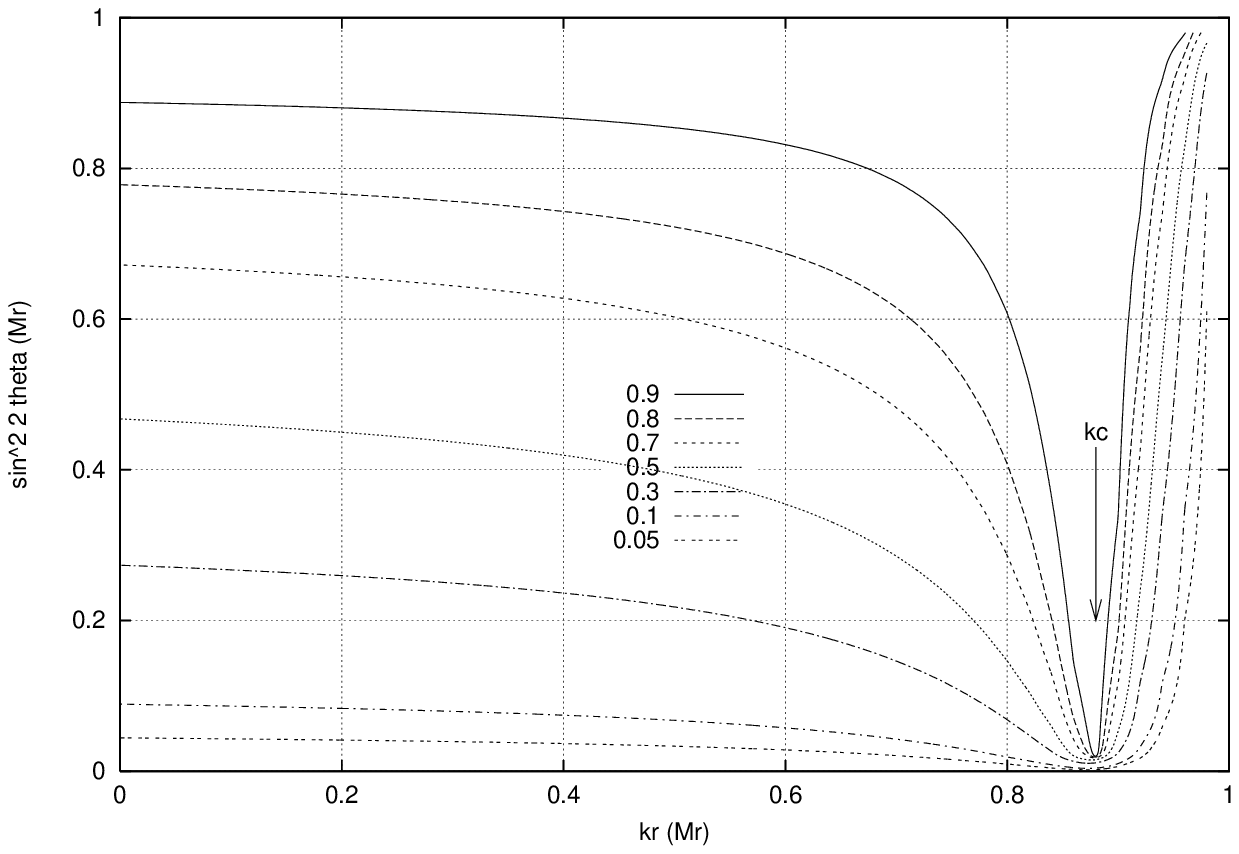}}\\
 Figure~\ref{fig:sin}(b): $\tan\beta=50$, ${\cal M_R}=10^{16}$ GeV
 \end{center}
 \vspace{-1.5em}
\caption{%
The contour plots of the mixing angle $\sin^2 2\theta_{23}$ 
 at the weak scale with $\tan \beta =50$.
We show two cases of
(a):${\cal M_R}=10^{14}$ GeV and
(b):$10^{16}$ GeV.
The horizontal axes show the value of $\kappa_r$
at $\mu ={\cal M_R}$ scale, 
and the vertical axes show 
the mixing angle at $\mu ={\cal M_R}$
The values of $\sin^2 2\theta_{23}$ at the weak scale 
are determined by inputting the initial values of 
$\sin^2 2\theta_{23}({\cal M_R})$ and 
$\kappa_r({\cal M_R})$. 
Around $\kappa_c$, the RGE effects make 
the weak scale mixing be large at any initial conditions.
}
 \label{fig:sin}
\end{figure}

Figures \ref{fig:sin} are 
the contour plots of the mixing angle $\sin^2 2\theta_{23}$ 
at the weak scale.
These are obtained by solving Eq.(\ref{eqn:RGEkappa2b})
 with the various $\kappa_r({\cal M_R})$(horizontal axis)
 and $\sin^2 2\theta_{23}({\cal M_R})$(vertical axis).
Here $\kappa_c$ in Figs.~\ref{fig:sin} 
 is the value at ${\cal M_R}$ that induces  
 $\kappa_r(\mbox{weak}) = 1$.
In the parameter region of 
 $\kappa_r({\cal M_R}) < \kappa_c$, 
 there are no significant RGE corrections  
 and then the mixing angle does not change drastically. 
The energy dependence of the mixing angle in this case
 is similar to the solid lines in Figs.~\ref{fig:EngDepOfSin}.
 Next in the case of $\kappa_r({\cal M_R}) \simeq \kappa_c$,
 which means $\kappa_r(\mbox{weak}) \simeq 1$,
 the mixing angle at the weak scale is strongly enhanced
 near the weak scale,
 as the dashed lines in Figs.~\ref{fig:EngDepOfSin}.
Then the mixing angle at the weak scale becomes maximal
 independently of the mixing angle at ${\cal M_R}$.
Finally in the case of $\kappa_r({\cal M_R}) > \kappa_c$,
 the mixing angle at the weak scale becomes small
 even for the large mixing angle at ${\cal M_R}$,
 where the energy dependence of the mixing angle is
 similar to the dotted lines in Figs.~\ref{fig:EngDepOfSin}.
{}From these arguments we can easily see that 
 there is 
 the large enhancement of the mixing angle from the RGE 
 around $\kappa_r \simeq 1$.

\par

Figures \ref{fig:m3} are
 the contour plots of the heaviest neutrino mass $m_3$ 
 at the weak scale.
The horizontal and vertical axes are the same as 
 those of Figs.~\ref{fig:sin}. 
We determine the masses of neutrino
 by substituting the parameters in \eqref{EVal2KrSin}
 with the result of Figs.~\ref{fig:sin}
 and the experimental value
 $\delta m^2_{23} \simeq 1.3 \times 10^{-3}\;\mbox{eV}^2$.
Since we use the experimental value of $\delta m^2_{23}$
 instead of evaluating the RGE of \eqref{RGEkappa2c},
 we can determine the masses without any additional input parameters.
As $m_3$ becomes large, the region 
 of $\kappa_r({\cal M_R})$ is limited
 around $\kappa_c$. 
Compared with Figs.~\ref{fig:sin},
 it is found
 that the region where the heaviest mass $m_3$ is larger 
 than $O(0.1\mbox{eV})$ corresponds to the region
 where the mixing angle at the weak scale 
 is always larger than 0.9
 despite the small mixing at ${\cal M_R}$ scale. 
%

\begin{figure}[htbp]
 \begin{center}
 \resizebox{.75\textwidth}{!}{\includegraphics{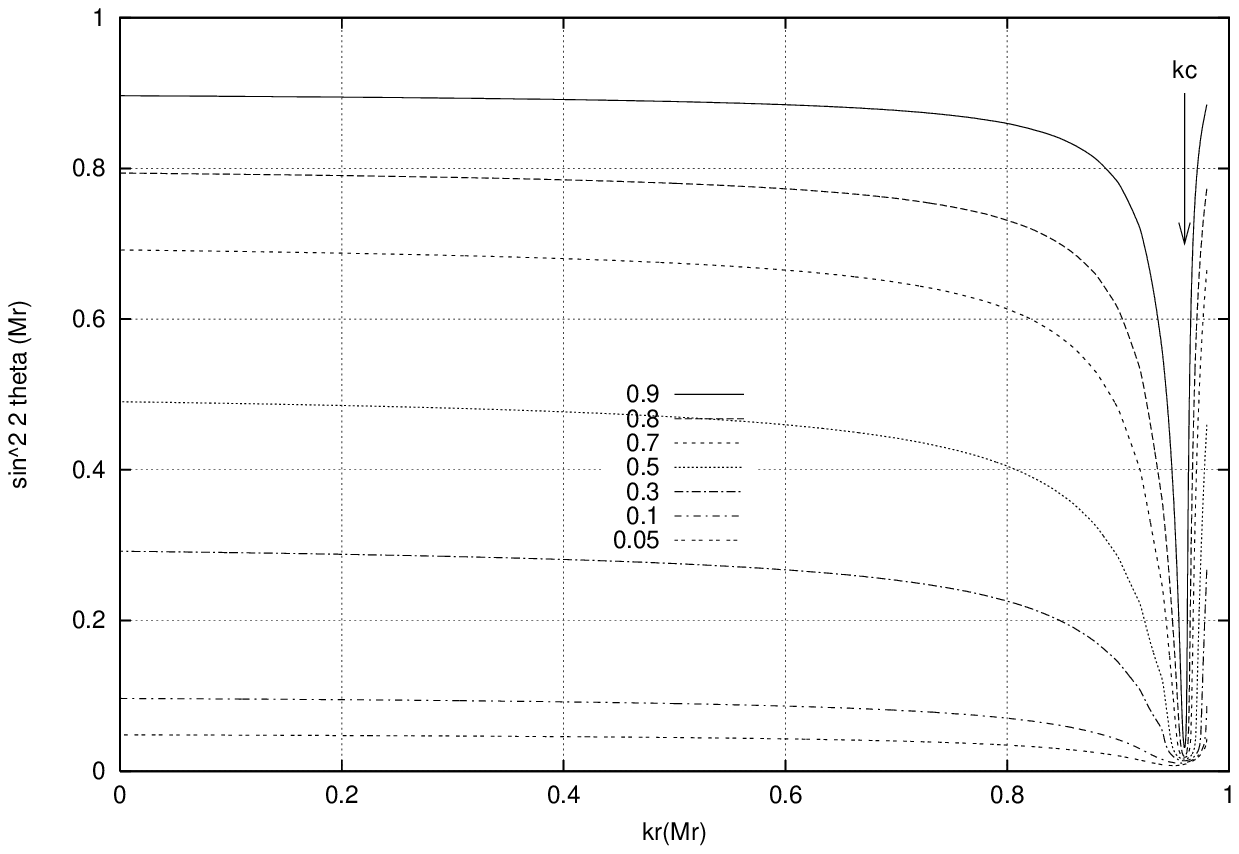}}\\
 Figure~\ref{fig:sin35}(a): $\tan\beta=35$, ${\cal M_R}=10^{14}$ GeV \\
 \vspace{.5em}\par\noindent
 \resizebox{.75\textwidth}{!}{\includegraphics{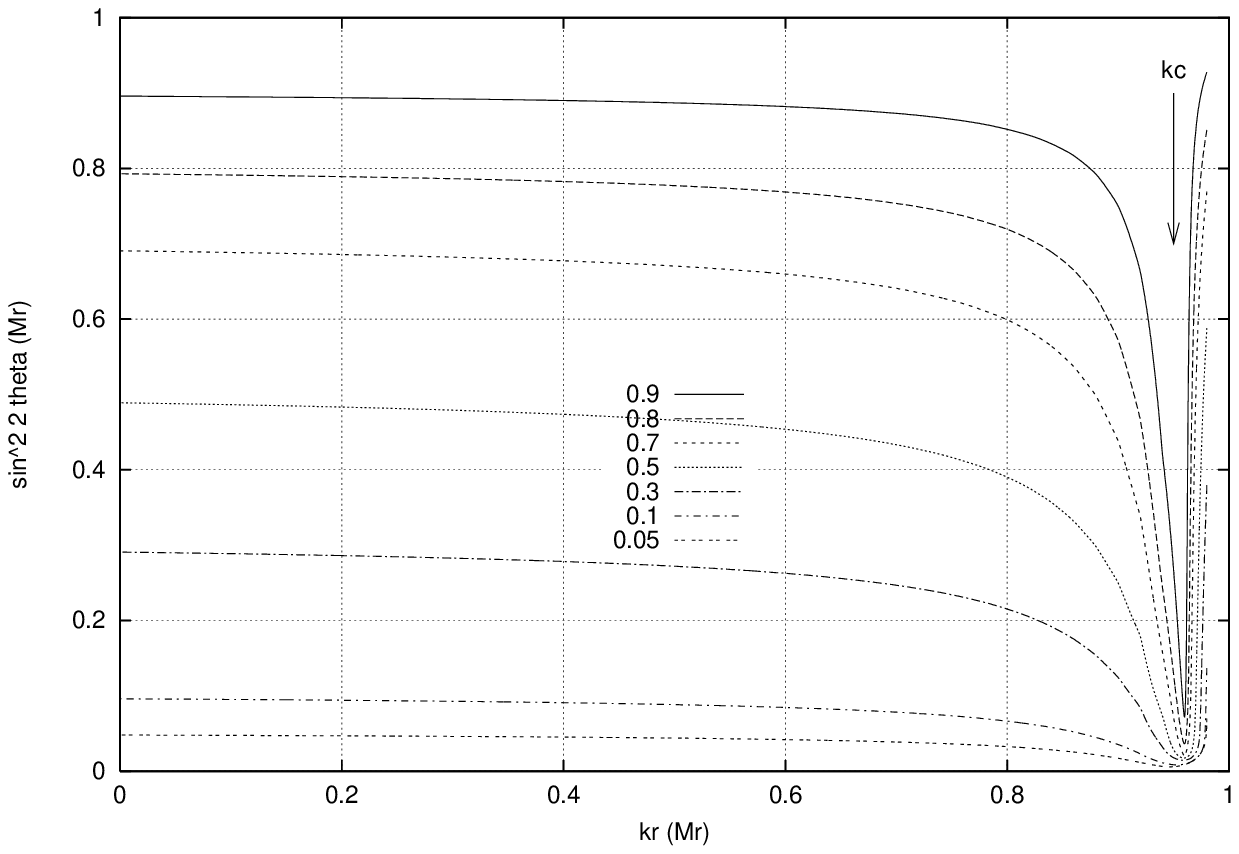}}\\
 Figure~\ref{fig:sin35}(b): $\tan\beta=35$, ${\cal M_R}=10^{16}$ GeV
 \end{center} 
 \vspace{-1.5em}
\caption{%
The contour plots of the mixing angle $\sin^2 2\theta_{23}$ 
 at the weak scale with $\tan \beta =35$.
In this case $\kappa_c$ is larger than that of Figs.~\ref{fig:sin}.
}
 \label{fig:sin35}
\end{figure}

Figures~\ref{fig:sin35} and Figures~\ref{fig:m335}
 correspond to Figures~\ref{fig:sin} and Figures~\ref{fig:m3}
 with another value of $\tan\beta=35$.
Compared with the case of $\tan\beta=50$,
 the value of $\kappa_c$ are just shifted to the right 
 in the case of $\tan\beta=35$.
Around $\kappa_r({\cal M_R}) \simeq \kappa_c$,
 the value of 
 $m_3$ is slightly larger than that of $\tan\beta=50$
 at the same value of $\sin^2 2\theta_{23}$.
%
%
In general,
 the smaller $\tan\beta$ becomes, 
 the more $\kappa_c$ approaches one, 
 and the value of $m_3$ around $\kappa_r({\cal M_R}) \simeq \kappa_c$
 becomes larger.
The region where 
 the maximal enhancement of the mixing angle is derived by 
 the RGE exists even in the case of 
 small $\tan \beta$. 
The value of $m_3$ around $\kappa_c$ becomes larger 
 corresponding to  
 the smaller value of $\tan\beta$.

We stress here that the large enhancement factor
 at the weak scale induced by
 $[(1+ \kappa_r)/(1- \kappa_r)]$ in \eqref{RGEkappa2b}
 is not the fine-tuning.
This factor must be inevitably large with 
$\kappa_r(\mbox{weak})\simeq 1$ if $m_2$ and $m_3$ are
 larger than  
 $O(0.1\mbox{eV})$. 
It is worth noting that
 the enough enhancement of the mixing angle
 can be obtained even in the case of 
 $O(0.1\mbox{eV})$ neutrino masses.
Even in the small $\tan \beta$, 
 $O(1\mbox{eV})$ mass is enough
 to obtain this RGE enhancement.

%
%
\section{Summary and Discussion}

In this paper we analyzed the RGE effects
of the neutrino flavor mixing between $\nu_\mu$ and $\nu_\tau$
in the MSSM
with right-handed neutrinos.
The experimental result of the Superkamiokande 
 suggests the 
 large neutrino flavor mixing between 
 $\nu_{\mu}$ and $\nu_{\tau}$ with 
 $\delta m^2_{23} \simeq 10^{-3}$ eV. 
Then we found that 
 the mixing angle between $\nu_{\mu}$ 
 and $\nu_{\tau}$  
 receives significant corrections
 from the RGE
 in the case of $m_{\nu_s} \approx m_{\nu_e} \ll 
  m_{\nu_{\mu}} \approx m_{\nu_{\tau}}$. 
In this mass spectrum, 
 the mixing angle at the decoupling 
 scale of right-handed neutrinos must be small 
 when $O(0.1\mbox{eV}) \leq m_{\nu_{\tau}} 
 (\approx m_{\nu_{\mu}})$. 

\begin{figure}[htbp]
 \begin{center}
 \resizebox{.75\textwidth}{!}{\includegraphics{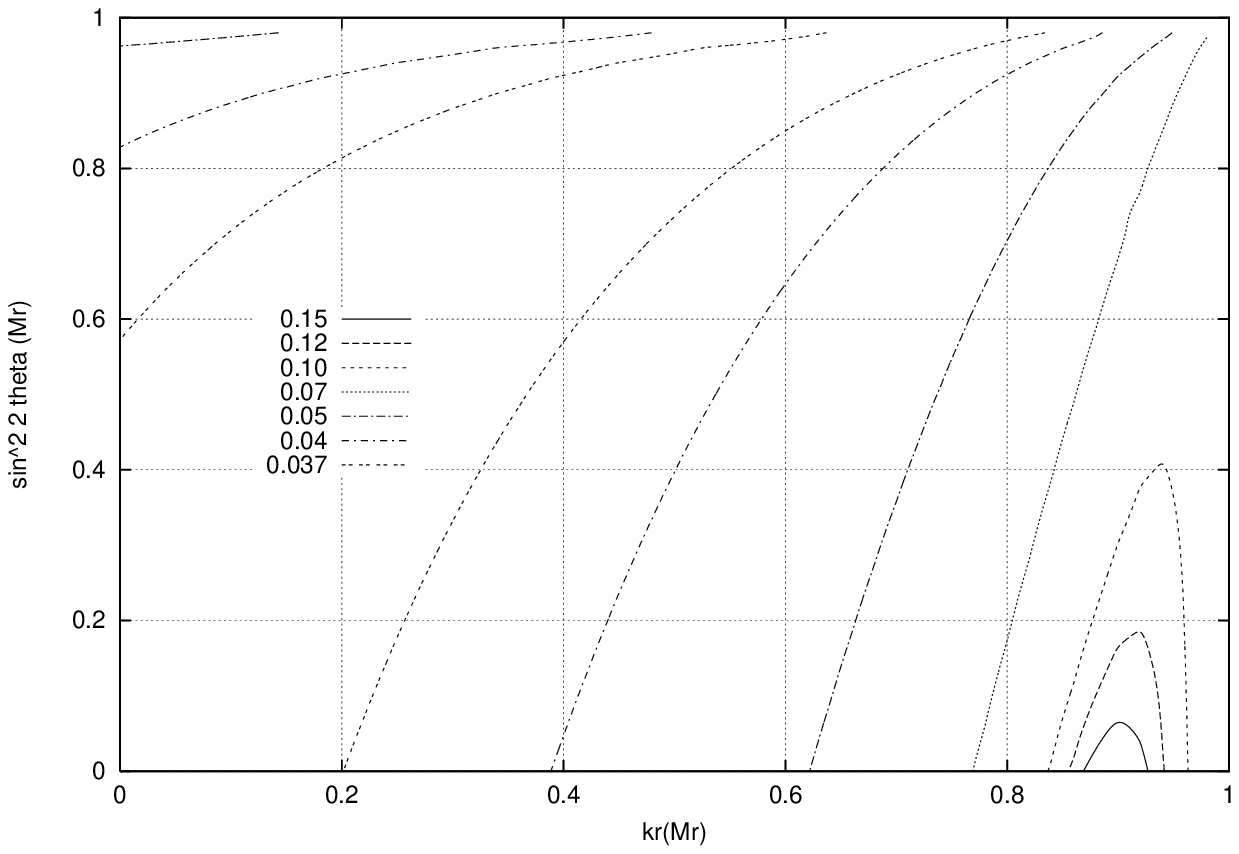}}\\
 Figure~\ref{fig:m3}(a): $\tan\beta=50$, ${\cal M_R}=10^{14}$ GeV \\
 \vspace{.5em}\par\noindent
 \resizebox{.75\textwidth}{!}{\includegraphics{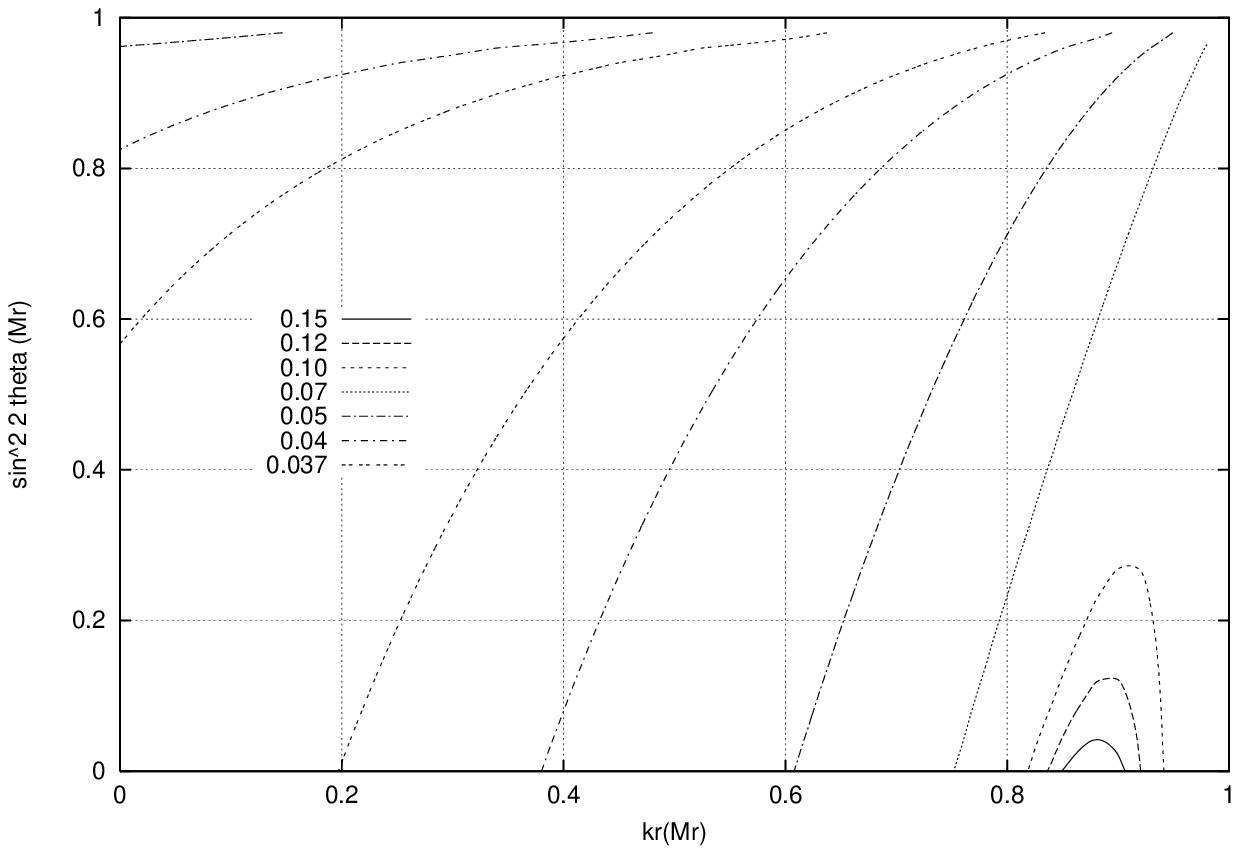}}\\
 Figure~\ref{fig:m3}(b): $\tan\beta=50$, ${\cal M_R}=10^{16}$ GeV
 \end{center}
 \vspace{-1.5em}
 \caption{
The contour plots of the heaviest neutrino mass $m_3$ 
at the weak scale with $\tan\beta = 50$.
The horizontal and vertical axes are the same as Figs.~\ref{fig:sin}.
The values of $m_3$ at the weak scale 
are determined by inputting the initial values of 
$\sin^2 2\theta_{23}({\cal M_R})$ 
and $\kappa_r({\cal M_R})$, and 
the experimental value of 
$\delta m_{23}^2 = 1.3 \times 10^{-3} {\rm eV}^2$. 
As the mass becomes heavy, 
the region of $\kappa_r({\cal M_R})$ is limited 
around $\kappa_c$.
}
 \label{fig:m3}
\end{figure}
\begin{figure}[htbp]
 \begin{center}
 \resizebox{.75\textwidth}{!}{\includegraphics{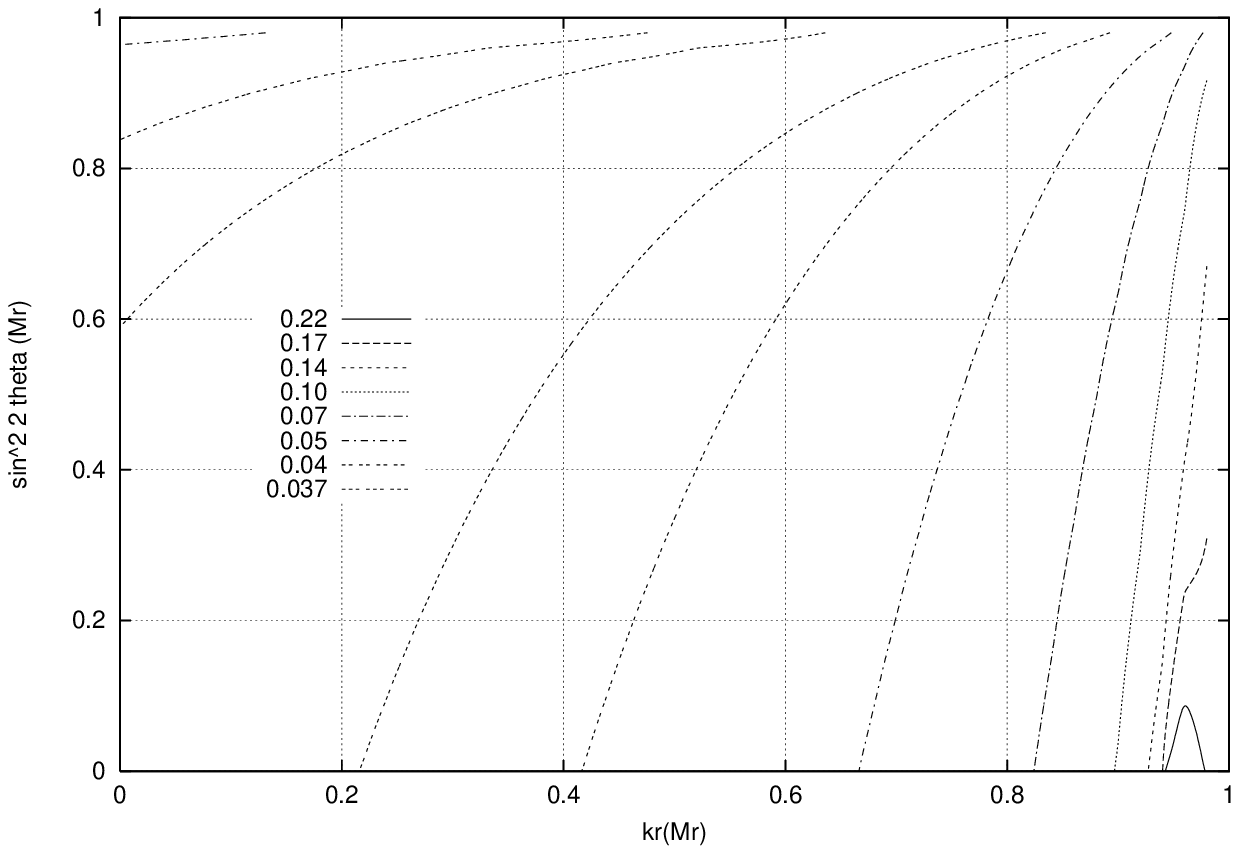}}\\
 Figure~\ref{fig:m335}(a): $\tan\beta=35$, ${\cal M_R}=10^{14}$ GeV \\
 \vspace{.5em}\par\noindent
 \resizebox{.75\textwidth}{!}{\includegraphics{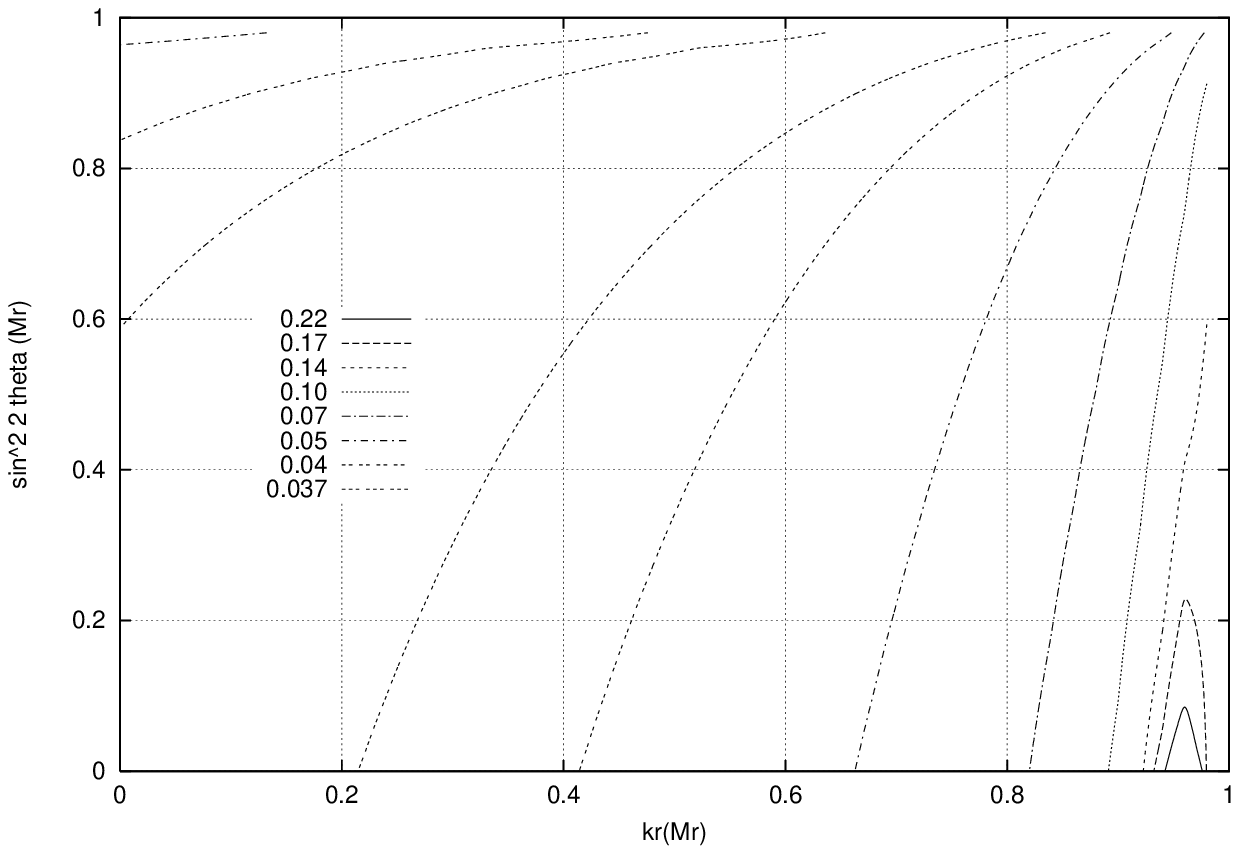}}\\
 Figure~\ref{fig:m335}(b): $\tan\beta=35$, ${\cal M_R}=10^{16}$ GeV
 \end{center} 
 \vspace{-1.5em}
\caption{%
The contour plots of the heaviest neutrino mass
 at the weak scale with $\tan \beta =35$.
The region where $m_3$ is larger than $O(0.1\mbox{eV})$
 corresponds to the region 
 in which the maximal mixing angle is realized 
 at low energy. 
}
 \label{fig:m335}
\end{figure}

Finally we give a brief comment 
 about the LSND experimental result. 
In the above mass spectrum, 
 both $m_{\nu_{\mu}}$ and $m_{\nu_{\tau}}$
 must be of order 1 eV
 to explain the LSND result.
In this case, the RGE analysis
 shows that
 the mixing angle at the decoupling 
 scale of right-handed neutrinos must be 
 negligibly small\footnote{
Although we have not shown the lines of 1 eV in 
 Figs.~\ref{fig:m3} and Figs.~\ref{fig:m335}, 
 there exist
 these lines. 
They are too small to be drawn apparently 
 in the large $\tan \beta$.
}.
{}From the view point of the model building,
 we must find the fundamental theory which induces
 the small mixing angle at the high energy scale
 in the four neutrino scenario 
 with the mass spectrum 
 of $m_{\nu_s} \approx m_{\nu_e} \ll 
  m_{\nu_{\mu}} \approx m_{\nu_{\tau}}$.

%
%
\vspace{2em}\par
\noindent
{\bf Acknowledgments}\par
We would like to thank M.~Tanimoto 
 for useful discussions. 
NO would like to thank
 K.~Hagiwara, J.~Hisano and Y.~Okada
 for useful discussions and comments.
NH would like to thank
 S.~Raby and K.~Tobe for helpful discussions and 
 comments. 
The work of NO is financially supported by the JSPS
 Research Fellowships for young scientist, No.2996.
NH is partially supposed by DOE grant DOE/ER/01545-753.

%

\newpage
%
%
%
\newpage
%
%
%

%
%
%
%
%
%

\end{document}